\documentclass[%
 reprint,
 amsmath,amssymb,
 aps,
prd,
longbibliography,
]{revtex4-2}

\usepackage{graphicx}
\usepackage{dcolumn}
\usepackage{bm}
\usepackage{hyperref} 
\usepackage[caption=false]{subfig}
\usepackage{bm}
\hypersetup{
	colorlinks = true,
    linkcolor = Black,
    urlcolor  = Black,
    citecolor = Black
}

\usepackage{verbatim}
\usepackage[amssymb]{SIunits}
\usepackage{tabularx}
\usepackage[dvipsnames]{xcolor}
\usepackage{tikz}
\usepackage{orcidlink}
\usepackage[normalem]{ulem}
\usepackage{soul}
\usepackage{balance}

\def\tr{\hbox{Tr}}

\begin{document}

\title{Generalized HMC using Nambu mechanics for lattice QCD}
\author{Erik Lundstrum}
\email{ell2156@columbia.edu}
\affiliation{Department of Physics, Columbia University, New York, NY 10027, USA}
\date{\today}

\begin{abstract}
    We describe a generalization of the hybrid Monte Carlo (HMC) algorithm in which the molecular dynamics (MD) steps utilize Nambu generalized Hamiltonian dynamics. Nambu mechanics contains multiple Hamiltonian functions. Using one of these Hamiltonians in the Metropolis accept/reject step is enough to guarantee the correct target distribution. The other Hamiltonians can include arbitrary functions of the gauge field. We examine the simplest case with two Hamiltonians and choose the second to include non-local functions of the gauge field. In this way, the links are updated with instantaneous knowledge of far-separated link variables. This represents a promising method for reducing critical slowing down in lattice QCD simulations.
\end{abstract}

\maketitle


\section{Introduction}

The main algorithm used in lattice QCD simulations is the Hybrid Monte Carlo (HMC) \cite{DUANE1987216}, in which all gauge links are updated in parallel via formulating the theory in terms of an evolving classical system. An issue with which LQCD simulations must contend is critical slowing down (CSD) when approaching the continuum limit. Despite the non-locality from dynamical fermions, the dominant force that enters the HMC molecular dynamics (MD) is from the local part of the QCD action which only couples nearby links. As such, it takes many classical trajectories for changes to diffuse across large distances on the lattice. This results in large autocorrelation times for observables which are sensitive to long-distance fluctuations in the gauge field \cite{PhysRevD.41.1953, Schaefer_2011}.

The strategy we present here relies on a modified MD based on Nambu mechanics \cite{nambu_mech} in which the forces from arbitrary non-local functions of the dynamical variables can be directly included in the MD evolution. The hope is that this additional non-locality may provide the mechanism for changes to more rapidly diffuse across a lattice and thus reduce CSD. Nambu mechanics is characterized by $d$ conjugate variables and $d-1$ conserved Hamiltonian functions that dictate the motion. Nambu mechanics preserves the volume of phase space and is exactly reversible, making it a natural candidate for an extended HMC algorithm. 

Here, we focus on the simplest generalization of Hamiltonian mechanics. In this case a dynamical degree of freedom is described by three real variables $(p,q,r)$ with their time evolution determined by two Hamiltonian functions $(H,G)$. In principle, this algorithm may be extended to arbitrary $d$. The main idea is that only one of the Hamiltonian functions $H$ is required for a Metropolis accept/reject step. This is enough to generate samples distributed according to the target distribution with few conditions imposed on choice of the other Hamiltonian $G$. This auxiliary Hamiltonian $G$ can be freely chosen to increase the sampling efficiency. Here, we suggest choosing $G$ to be a function of a non-local observable. By doing so, the forces felt by the gauge links have components directed along the gradients of this second, non-local function, thereby providing the customizable, long-range communication lacking in the usual HMC. Other than this modification to the MD steps, the algorithm proceeds in the same way as the HMC. 

The goal of this paper is to present this new framework and demonstrate its correctness. To do so, we provide a mathematical proof of detailed balance and perform a numerical test in the physically realistic environment of 4D pure $SU(3)$ gauge theory. High-precision plaquette measurements are found to be consistent with the usual HMC. We also present preliminary tests of plaquette and Wilson loop autocorrelations that, while not aimed at providing definitive evidence of mitigating CSD, demonstrate that the auxiliary Hamiltonian can influence the sampling efficiency of observables. We believe that this result is promising and motivates further study.

The organization of this paper is as follows. Section \ref{HMC} reviews the HMC algorithm and Section \ref{nambu_mech} describes how it may be extended to Nambu mechanics. The following Section \ref{qcd} discusses how the HMC with Nambu mechanics may be applied to lattice gauge theory. In Section \ref{numerics} we perform numerical tests in pure SU(3) gauge theory and find that the Nambu HMC produces results consistent with a correct algorithm. We also make preliminary investigations of how the Nambu HMC affects sampling efficiency. In the final section, we discuss the results and present conclusions.

\section{HMC}
\label{HMC}

Our objective is to find the expectation value of an observable $A(q)$ where the dynamical field $q$ is governed by the action $S(q)$ 
\begin{equation}
    \langle A \rangle = \frac{1}{Z}\int [dq] A(q) \exp\left( - S(q)\right),
\end{equation}
\noindent with the partition function
\begin{equation}
    Z = \int [dq] \exp(-S(q)).
\end{equation}
\noindent We estimate $\langle A \rangle$ with a stochastic process in which a set of field configurations $\{q\}$ are generated at random with probability $P(q) = Z^{-1} e^{-S(q)}$. For a large number of samples $N$ the observable of interest can be calculated as
\begin{equation}
    \langle A \rangle = \frac{1}{N} \sum_{k=1}^N A(q_{(k)}) + O(1/\sqrt{N}),
\end{equation}
\noindent where the final term $O(1/\sqrt{N})$ is the statistical error associated with the estimate. To accomplish this, we use a Markov chain where the sample $q'$ is generated from a previous state of the system $q$ with a probability of transition $P_T(q \rightarrow q')$. If the Markov process is ergodic and satisfies the detailed balance condition
\begin{equation}
    P(q) P_T(q \rightarrow q') = P(q') P_T(q' \rightarrow q),
\end{equation}
\noindent the algorithm will correctly produce samples according to the desired probability distribution $P(q)$. The procedure used to generate the new samples ideally has a high acceptance rate and minimizes correlations between samples.

The HMC accomplishes this by evolving the field $q$ in a fictitious ``computer" time $t$ using Hamiltonian mechanics \cite{DUANE1986143}. We treat the field variables $q$ as the coordinates of this classical system and generate fictitious momenta to complete the phase space $(p,q)$. We choose the Hamiltonian
\begin{equation}
    H(p,q) = \frac{1}{2} p^2 + S(q).
\end{equation}
\noindent The field $q$ is updated by using Hamilton's equations
\begin{equation}
\begin{aligned}
    & \dot{q} = p, \\
    & \dot{p} = - \frac{\partial S}{\partial q}. \\
\end{aligned}
\label{momenta_hamiltonian_update}
\end{equation}
\noindent The procedure is to generate the initial momentum $p(0)$ at random from a Gaussian distribution $\exp(-p^2/2)$ and use the initial coordinates $q(0) = q$. The new field configuration is produced by integrating Hamilton's equations for trajectory of length $t$:  $q(t)=q'$. The trajectory can be obtained by alternatively updating the coordinates and momentum for a discrete time increment $\tau$. At the end of the trajectory the new configuration is accepted with probability $P_A = \text{min}(1, e^{-\Delta H})$ where $\Delta H = H(p(t),q(t)) - H(p(0),q(0))$. For this scheme to satisfy detailed balance the integrator for Hamilton's equations must be reversible and preserve phase space volume. The simplest choice satisfying these criteria is the leapfrog
\begin{equation}
\begin{aligned}
    &p^{n+1/2} = p^{n} - \frac{\tau}{2} \left.\frac{\partial S}{\partial q}\right|_{q^n} , \\
    &q^{n+1} = q^n + \tau p^{n+1/2}, \\
    &p^{n+1} = p^{n+1/2} - \frac{\tau}{2} \left.\frac{\partial S}{\partial q}\right|_{q^{n+1}}.
\end{aligned}
\label{HMC_leapfrog}
\end{equation}
\noindent Here, the superscript $n$ denotes the position in discretized computer time. To complete a trajectory of total length $t$, one performs $N = t/\tau$ iterations of the leapfrog.

More generally, this process works because we can supplement the phase space with the additional fictitious variables $p$ that don't enter the observables without changing the physical content of the theory. This is because performing the $p$ portion of the functional integral
\begin{equation}
    \int [dp][dq] A(q) \exp\left(-S(q) - p^2/ 2 \right),
\end{equation}
\noindent introduces a constant factor that cancels in the normalization  \cite{PhysRevLett.49.613}. 

\section{Nambu Mechanics}
\label{nambu_mech}

This section describes Nambu mechanics, emphasizing the features relevant to the algorithm. For a single degree of freedom, Nambu mechanics is defined on a $d-$dimensional phase space by $d-1$ conserved Hamiltonian functions. Familiar Hamiltonian mechanics corresponds to $d=2$. For $d=3$, the dynamical variables form a canonical triplet $(p,q,r)$ and the time evolution is dictated by the Hamiltonians $H$ and $G$. The time derivative of any function of the canonical variables $F(p,q,r)$ is expressed by the Jacobian
\begin{equation}
    \dot{F} = \frac{\partial(F,H,G)}{\partial(q,p,r)} = \sum_{a,b,c}\epsilon_{abc} \frac{\partial F}{\partial x_a} \frac{\partial H}{\partial x_b} \frac{\partial G}{\partial x_c}, 
\end{equation}
\noindent where in the final equality $\epsilon_{abc}$ is the antisymmetric tensor and the variables $x_a$ with $a = 1,2,3$ correspond to the canonical variables $q,p,r$, respectively. This may be generalized to a system of $N$ canonical triplets as
\begin{equation}
    \dot{F} = \sum_{i=1}^N \frac{\partial(F,H,G)}{\partial(q_i,p_i,r_i)}.
\label{Nambu_bracket}
\end{equation}
\noindent By the asymmetry of the Jacobian it is clear $\dot{H}=\dot{G}=0$. To satisfy the reversibility requirement for the algorithm, we restrict ourselves to Hamiltonians that satisfy $H(-p,q,r) = H(p,q,r)$ and $G(-p,q,r) = G(p,q,r)$. In this case, the time evolution is reversed by reflecting the set of variables $\{p_i\} \rightarrow \{-p_i\}$. Nambu mechanics also preserves the volume of phase space \cite{nambu_mech}, which together with reversibility means that it is suitable for an extended HMC algorithm.

To adapt the HMC algorithm to use Nambu mechanics we introduce an additional fictitious variable to the path integral
\begin{equation}
    \frac{1}{Z} \int [dp] [dr][dq] A(q) \exp\left(-S(q) - p^2 / 2 - r^2 / 2 \right),
\end{equation}
\noindent and interpret the sum $H(p,q,r) = S(q) + p^2/2 + r^2/2$ as one of the Hamiltonians of the Nambu system. We can now use the classical evolution equation in Eq.~\eqref{Nambu_bracket} to update the field variables $q$. At the beginning of a trajectory, $p$ and $r$ are drawn from a Gaussian distribution of unit variance, which along with the field variables $q$ form the initial state of the system. At the end of the trajectory we accept the new configuration $q(t)$ with probability $P_A = \text{min}(1, e^{-\Delta H})$.

\section{Application to lattice gauge theory}
\label{qcd}

In this section we apply Nambu mechanics to lattice gauge theory and describe the discrete updates needed to realize a Nambu mechanics version of the HMC. We show that an algorithm satisfying the detailed balance condition can be constructed with an auxiliary Hamiltonian $G$ that contains arbitrary functions of the gauge field variables.

Here, we replace the general classical coordinates $q$ with the SU($N$) matrices $U(x,\mu)$ of lattice gauge theory associated with the links joining nearest-neighbor sites in a hypercubic lattice. These gauge links are labeled by their lattice location $x$ and Euclidean direction $1 \leq \mu \leq 4$. We focus first on a single link $U$. The $SU(N)$ matrix $U$ is a constrained variable and, when needed, $U$ is viewed as a function of the $N^2 - 1$ real variables $q_a$ that parameterize the Lie group in the vicinity of a constant group element $U$ as
\begin{equation}
    U' = \exp \left(-\sum_{a=1}^{N^2-1} q_a T_a \right) U,
\label{parameterization_su3_group}
\end{equation}
\noindent where $T_a$ are the anti-hermitian Lie group generators and $1\leq a \leq N^2-1$ are adjoint representation indices. This allows us to write down derivatives with respect to the link variables, defined abstractly as $\boldsymbol{e}_a U$
\begin{equation}
    \boldsymbol{e}_a U = \left.\frac{\partial U'}{\partial q_a} \right|_{q=0} = -T_a U.
\label{derivative_field_space}
\end{equation}
\noindent Here, the minus sign is present so that differential operators satisfy $[\boldsymbol{e}_i, \boldsymbol{e}_j] = c^k_{ij} \boldsymbol{e}_k$ \cite{Kennedy_2013}. Derivatives of functions of $U$ are given by 
\begin{equation}
    \boldsymbol{e}_a S(U) = \left.\frac{\partial S(U')}{\partial q_a} \right|_{q=0}.
\label{derivative_function_U}
\end{equation}
We define the Nambu classical phase space for a single gauge link by assigning each $q_a$ in Eq.~\eqref{parameterization_su3_group} to a canonical triplet $(p_a, q_a,r_a)$. This defines the classical phase space of a single link: $\big\{U, \{p_a, r_a\}_{1\leq a \leq N^2-1}\big\}$. 

We focus on separable Hamiltonian functions of the form
\begin{equation}
\begin{aligned}
    &H(p,U,r) = \sum_{a=1}^{N^2-1}\frac{p_a^2}{2} + \sum_{a=1}^{N^2-1}\frac{r_a^2}{2} + S(U), \\
    &G(p,U,r) = g_1(p) + g_2 (r) + g_3(U). \\
\end{aligned}
\label{separable hamiltonians}
\end{equation} 
In accordance with the previous section, the function $S(U)$ entering $H$ is the target action we seek to simulate. Reversibility requires $g_1(-p) = g_1(p)$, but no restrictions are placed on the functions $g_2(r)$ or $g_3(U)$. The continuum trajectory defined by the Hamiltonians in Eq.~\eqref{separable hamiltonians} can be approximated by taking discrete steps of size $\tau$. This is done with staggered updates where one variable is evolved for a discrete time increment while the others are held constant.  

To find the form of the gauge link update we calculate its rate of change using Eqs.~\eqref{Nambu_bracket} and \eqref{derivative_field_space} and find
\begin{equation}
\begin{aligned}
    \dot{U} &= \sum_{a=1}^{N^2-1} \frac{\partial(H,G)}{\partial (p_a, r_a)} \boldsymbol{e}_a U \\
    &= \sum_{a=1}^{N^2-1} \frac{\partial(H,G)}{\partial (p_a, r_a)} \left( -T_a \right) U. \\
\end{aligned}
\end{equation}
\noindent For separable Hamiltonians $H$ and $G$, the factor multiplying $U$ independent of $U$, so the differential equation has the solution
\begin{equation}
\begin{aligned}
    U^{n+1}= \exp\left( -\tau \sum_{a=1}^{N^2-1} \frac{\partial(H,G)}{\partial (p_a, r_a)} T_a \right) U^n.
\end{aligned}
\label{link_update}
\end{equation}
\noindent Here, the superscript $n$ labels the position in the discretized computer time. This is a left-multiplication of the group element $U$ by an SU($N$) matrix, under which the Haar measure is invariant. Thus, this update conserves the phase space volume.

To find the $p_a$ and $r_a$ updates, first we calculate the rates of change $\dot{p}_a$ and $\dot{r}_a$ with Eq.~\eqref{Nambu_bracket}, everywhere making the replacement $\partial/\partial q_a \rightarrow \boldsymbol{e}_a$ and using Eq.~\eqref{derivative_function_U} for functions of the gauge links. Since $p_a$ and $r_a$ are unconstrained variables, their updates are linear
\begin{equation}
\begin{aligned}
    & p^{n+1}_a = p^n_a + \tau \bigg[\boldsymbol{e}_a G \cdot \frac{\partial H}{\partial r_a} - \boldsymbol{e}_a H \cdot \frac{\partial G}{\partial r_a}\bigg], \\
    & r^{n+1}_a = r^n_a + \tau \bigg[\boldsymbol{e}_a H \cdot \frac{\partial G}{\partial p_a} - \boldsymbol{e}_a G \cdot \frac{\partial H}{\partial p_a}\bigg]. \\
\end{aligned}
\label{real-values_updates}
\end{equation}
Here there is no summation over the repeated index $a$. The updates in Eqs.~\eqref{link_update} and \eqref{real-values_updates} must be combined into a reversible integrator. This is accomplished by symmetrizing the updates. One possible option is a $PRURP$ scheme
\begin{equation}
\begin{aligned}
    & p^{n+1/2}_a = p^n_a + \frac{\tau}{2} \bigg[\boldsymbol{e}_a G \cdot \frac{\partial H}{\partial r_a} - \boldsymbol{e}_a H \cdot \frac{\partial G}{\partial r_a}\bigg]_{U^n, r^n},\\
    & r^{n+1/2}_a = r^n_a + \frac{\tau}{2} \bigg[\boldsymbol{e}_a H \cdot \frac{\partial G}{\partial p_a} - \boldsymbol{e}_a G \cdot \frac{\partial H}{\partial p_a}\bigg]_{U^n, p^{n+1/2}},\\
    & U^{n+1}= \exp\left( -\tau \sum_{a=1}^{N^2-1} \frac{\partial(H,G)}{\partial (p_a, r_a)} T_a \right) U^n \bigg|_{p^{n+1/2}, r^{n+1/2}},\\
    & r^{n+1}_a = r^{n+1/2}_a + \frac{\tau}{2} \bigg[\boldsymbol{e}_a H \cdot \frac{\partial G}{\partial p_a} - \boldsymbol{e}_a G \cdot \frac{\partial H}{\partial p_a}\bigg]_{U^{n+1}, p^{n+1/2}}, \\
    & p^{n+1}_a = p^{n+1/2}_a + \frac{\tau}{2} \bigg[\boldsymbol{e}_a G \cdot \frac{\partial H}{\partial r_a} - \boldsymbol{e}_a H \cdot \frac{\partial G}{\partial r_a}\bigg]_{U^{n+1}, r^{n+1}}.
\end{aligned}
\label{PRUPR_scheme}
\end{equation}
During the $p$ steps above, the entire set of variables $\{p_a^n\} \rightarrow \{p_a^{n+1/2}\}_{1\leq a \leq N^2-1}$ are updated in parallel. The same is true for the $r$ steps. A trajectory with total length $t$ consists of $N = t/\tau$ iterations of the above. The reversibility of this scheme dictates that the first finite-step-size errors in the conservation of $H$ and $G$ should occur at order $\tau^2$ \cite{leimkuhler_reich_2005}.

To generalize this to the full lattice of link variables, we assign Nambu canonical triplets to each link $U(x,\mu)$ in the manner described above. The total classical phase space is $\big\{ U(x,\mu), \{p_a(x,\mu), r_a(x,\mu)\}_{1\leq a\leq N^2-1}\big\}$. Derivatives which enter the classical evolution equations are extended to
\begin{equation}
    \frac{\partial}{\partial p_a} \rightarrow \frac{\partial}{\partial p_a^{x,\mu}}, \hspace{0.2cm} \frac{\partial}{\partial r_a} \rightarrow \frac{\partial}{\partial r_a^{x,\mu}}, \hspace{0.2cm} \boldsymbol{e}_a \rightarrow \boldsymbol{e}_a^{x,\mu},
\end{equation}
where each of these return zero when acting on a variable with different values of $x$ or $\mu$. The Hamiltonian H in Eq.~\eqref{separable hamiltonians} now includes the sums over lattice locations and directions
\begin{equation}
    H(p,U,r) = \sum_{x,\mu} \sum_{a=1}^{N^2-1}\frac{p_a^2(x,\mu)}{2} + \sum_{x,\mu} \sum_{a=1}^{N^2-1}\frac{r_a^2(x,\mu)}{2} + S(U).
\label{many_link_ham}
\end{equation}
Each step of the update scheme in Eq.~\eqref{PRUPR_scheme} is performed in parallel for every variable of that type on the lattice, i.e. the entire set $\{p_a^n(x,\mu)\rightarrow p_a^{n+1/2}(x,\mu)\}$ is updated during the first line.

Since the updating scheme remains reversible and volume-preserving, this algorithm will satisfy detailed balance. A proof of this can be found in Appendix \ref{detailed_balance}. This is true for \emph{any} form of the auxiliary Hamiltonian $G$, provided it is separable and satisfies the reversibility condition. This means that $G$ can contain non-local functions of the gauge links or something more exotic. Dynamical fermions can be included in this algorithm in the usual way by adding them to the target action $S(U)$. It should be noted that the forces from functions of the gauge field need not be reevaluated between adjacent $p$ and $r$ updates, meaning that this algorithm does not require additional fermion force evaluations beyond those present in the usual HMC.

It is interesting to note the compatibility of these updates with the gauge-invariance of the action $S(U)$. For the standard HMC the fictitious momenta and their rates of change transform in the adjoint representation under gauge transformations. In this case, the gauge symmetry manifests itself in an indifference as to whether a gauge transformation is made before or after a MD update \cite{DUANE1986143}. The transformation properties of the Nambu evolution updates can be studied by expanding the Nambu bracket appearing in Eq.~\eqref{link_update}
\begin{equation}
    \dot{U}(x,\mu) U(x,\mu)^{-1} = - \sum_{a=1}^{N^2-1}\bigg[ \frac{\partial H}{\partial p_a^{x,\mu}} \frac{\partial G}{\partial r_a^{x,\mu}} - \frac{\partial H}{\partial r_a^{x,\mu}} \frac{\partial G}{\partial p_a^{x,\mu}} \bigg]T_a.
\end{equation}
\noindent Here, there is no summation over the repeated indices $(x,\mu)$. The presence of two adjoint indices in the brackets on the right makes it clear that the rates of change of the phase space variables and the phase space variables themselves can't simultaneously transform in the adjoint representation under gauge transformations. Thus, for a generic choice of function $G$, the updates cannot be consistent with gauge symmetry in the same way as the standard HMC. Regardless, the algorithm preserves the Haar measure and has the gauge-invariant statistical weight as a fixed point. As such, it still functions as a correct algorithm for use in lattice QCD. This is confirmed by the numerical experiments that follow.

\section{Numerical tests}
\label{numerics}

It is prudent to test the Nambu HMC in a non-trivial case to verify that the algorithm behaves as expected. We discuss two issues: the correctness of the algorithm and the sampling efficiency. We simulate four-dimensional SU($3$) pure gauge theory on an $8^4$ lattice with periodic boundary conditions using the Wilson gauge action. This is done for two choices of auxiliary Hamiltonian containing non-local functions of the gauge links. 

The Wilson gauge action enters the main Hamiltonian $H$ in Eq.\eqref{many_link_ham} and is given by
\begin{equation}
    S(U) = \frac{\beta}{3} {\rm Re} \sum_{x, \nu>\mu} \tr (1 - P_{\mu\nu}(x)),
\end{equation}
\noindent where the plaquette $P_{\mu\nu}$($x$) is the ordered product of links around the  $1\times 1$ square with corner at $x$ and oriented in the $(\mu,\nu)$-plane
\begin{equation}
    P_{\mu\nu}(x) = U(x, \mu) U(x + \hat{\mu}, \nu) U(x + \hat{\nu}, \mu)^\dag U(x,\nu)^\dag.
\end{equation}
\noindent Here, $\hat{\mu}$ indicates the unit vector in the $\mu$ direction.

\begin{table*}[]
\begin{center}
\begin{tabular}{ |p{0.6cm}||p{2.cm}|p{1.0cm}|p{1.0cm}|p{1.cm}|  }
 \hline
 \multicolumn{5}{|c|}{NHMC PL Plaquette values $(8^4)$} \\
 \hline
 $\beta$ & plaquette & MD steps & accpt. rate & trajs\\
 \hline
 1.0 & $0.939857(73)$ & 10  & 0.70 & 2k \\
 3.0 & $0.794987(32)$  & 25  & 0.80 & 10k \\
 5.6 &  $0.475512(75)$& 45 & 0.84 & 100k  \\
 7.0 &  $0.328344(23)$ &  50 &  0.79 & 20k \\
 10.0&  $0.216650(17)$ & 65& 0.80 & 20k \\
 \hline
\end{tabular}
\begin{tabular}{ |p{0.6cm}||p{2.cm}|p{1.0cm}|p{1.0cm}|p{1.cm}|  }
 \hline
 \multicolumn{5}{|c|}{HMC Plaquette values $(8^4)$} \\
 \hline
 $\beta$ & plaquette & MD steps & accpt. rate & trajs\\
 \hline
 1.0  & $0.939913(55)$   & 25 & 0.79 & 2k \\
 3.0  &  $0.794994(30)$  & 45 & 0.80 & 10k \\
 5.6  & $0.475446(72)$ & 60 & 0.80 &  100k \\
 7.0  & $0.328291(48)$   & 80 & 0.77 & 20k\\
 10.0 & $0.216656(16)$ & 100 &  0.77& 20k \\
 \hline
\end{tabular}
\end{center}
\caption{High precision comparison of the Wilson action per plaquette for the HMC and Nambu HMC with Polyakov loops (NHMC PL). Statistical errors are calculated using the jacknife method and are stated in parentheses. Listed along with the measurements are the MD steps per trajectory and acceptance rates.}
\label{plaq_values_comparison}
\end{table*}

For the test of correctness we perform high-precision plaquette measurements for five values of $\beta$ and compare them to the usual HMC. For this test we choose an auxiliary Hamiltonian $G$ that is quadratic in $r_a(x,\mu)$ and is a function of Polyakov loops. The Polyakov loop $L(x,\mu)$ is the product of links wrapping around the lattice of side length $N$ and back to the origin
\begin{equation}
    L(x,\mu) = \prod_{i=0}^{N-1} U(x + i\cdot\hat{\mu}, \mu).
\end{equation}
The auxiliary Hamiltonian for this test is
\begin{equation}
    G(U,r) = \gamma\sum_{x,\mu}\sum_{a=1}^{8} \frac{r_a^2(x,\mu)}{2} + \frac{\kappa}{3}  {\rm Re} \sum_{\text{indpt.} (x,\mu)} \tr L(x,\mu),
\label{aux_ham_poly_loops}
\end{equation}
\noindent where the summation is performed over independent Polyakov loops. We arbitrarily choose the parameters in the auxiliary Hamiltonian as $\gamma = \kappa = 0.5$. The results of this test are given in Table \ref{plaq_values_comparison}. The Nambu HMC uses the integrator given in Eq.~\eqref{PRUPR_scheme} and the HMC uses a standard leapfrog. The trajectory length is $t=2$ for all tests performed. We find that all plaquette values agree within statistical error, verifying the correctness of the Nambu HMC.

To compare the sampling efficiency of the algorithms we need a cost metric by which to compare them. We opt for a simple comparison of the number of Wilson gauge force evaluations. This is done with the view that in realistic simulations fermions will dominate the cost per-update, so the additional cost from evaluating the gradients of the non-local functions of the gauge links won't add a large overhead.

To test the sampling efficiency we measure the normalized autocorrelation of observables. For a set of measurements $\{O\}$ the autocorrelation function is
\begin{equation}
    R_{\text{AC}} (k) = \frac{1}{N-k} \sum_{i=0}^{N-k-1} \Big( O(i) - \overline{O} \Big) \Big( O(i + k) - \overline{O} \Big),
\end{equation}
where $\overline{O}$ is the mean of the samples. The normalized autocorrelation is defined as
\begin{equation}
    \rho_{\text{AC}} (k) = \frac{R_{\text{AC}} (k)}{R_{\text{AC}} (0)}.
\end{equation}
In general, the decrease in autocorrelation will depend on the trajectory length. The trajectory length and step sizes are tunable parameters which likely take different optimal values in the Nambu HMC and HMC. The Nambu HMC contains additional free parameters available for tuning. In this case we are making preliminary investigations of the sampling efficiency and aren't making a full attempt to reduce CSD, so we choose the step sizes for the algorithms to have similar acceptance rates and use trajectories composed of a fixed number of Wilson force evaluations.

We perform three separate tests comparing the Nambu HMC and HMC, plotting the decrease of the normalized autocorrelation with increasing number of gradient evaluations for both the plaquette and the $3\times3$ Wilson loop. For these tests we choose the auxiliary Hamiltonian to be linear in $r_a(x,\mu)$. We also choose it to be a function of $3\times3$ Wilson loops to observe how adding a non-local function to the auxiliary Hamiltonian effects its sampling efficiency. Though used here for initial tests, we have no reason to expect that a large Wilson loop provides the optimal large-distance communication. An $N\times N$ Wilson loop $W_{\mu\nu}(N, x)$ is the ordered product of links around an $N\times N$ square with corner at $x$ and oriented in the $(\mu,\nu)$-plane
\begin{equation}
\begin{aligned}
    W_{\mu\nu}(N, x) = & \left[ \prod_{i=0}^{N-1} U(x + i\cdot \hat{\mu}, \mu) \right] \\ 
    & \left[\prod_{j=0}^{N-1} U(x + N \cdot \hat{\mu} + j \cdot \hat{\nu}, \nu)\right] \\
    & \left[ \prod^{N-1}_{k=0} U(x + N \cdot \hat{\nu} + k \cdot \hat{\mu}, \mu)\right]^\dag \\ & \left[ \prod_{l=0}^{N-1} U(x + l \cdot \hat{\nu}, \nu)\right]^\dag. \\
\end{aligned}
\end{equation}
The auxiliary Hamiltonian is
\begin{equation}
    G(U,r) = \gamma\sum_{x,\mu}\sum_{a=1}^{8} r_a(x,\mu) -  \frac{\kappa}{3} {\rm Re} \sum_{x, \nu>\mu} \tr \left( 1 - W_{\mu\nu}(3, x) \right).
\label{wilson_loops_function}
\end{equation}
We arbitrarily choose the parameters entering the auxiliary Hamiltonian as $\gamma = 1.0$ and $\kappa = 3.0$. For this auxiliary Hamiltonian, the choice $\gamma = 1.0$ and $\kappa = 0$ reduces the classical evolution equations to those of the usual Hamiltonian mechanics. Thus, this choice of auxiliary Hamiltonian can be considered a ``minimal deformation'' of the HMC.

\begin{figure}[]
    \centering
    \includegraphics[width=\columnwidth]{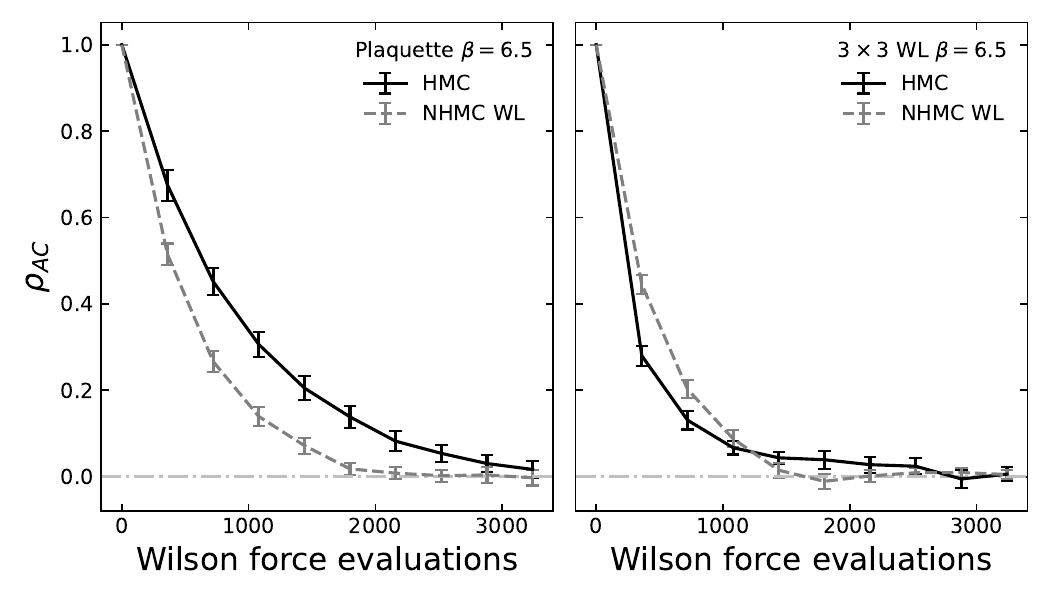}
    \captionsetup{skip=0pt}
    \caption{Comparison at $\beta = 6.5$ of the normalized autocorrelation $\rho_{\text{AC}}$ decreasing as a function of the number of Wilson force evaluations for the HMC and Nambu HMC with $3\times 3$ Wilson loops (NHMC WL). The left and right plots are the autocorrelation of the plaquette and $3\times3$ Wilson loop, respectively.}
    \label{b65_linG_WL}
\end{figure}

Test one is performed at a weak coupling $\beta = 6.5$, which is above the confining phase transition. The results are presented in Fig.~\ref{b65_linG_WL}. In this test there are 360 Wilson force evaluations per trajectory and autocorrelations are measured on 5000 trajectories. For the HMC, this corresponds to a total trajectory length of $t=7.2$ MD time units with acceptance rate $0.868$ and for the Nambu HMC, a trajectory of length $t=5.4$ MD units with acceptance rate $0.818$. The plotted errors in this test and those that follow are calculated using the jackknife method. We find that, compared to the HMC, the Nambu HMC more rapidly decorrelates the plaquette while producing a marginally slower decorrelation for the Wilson loop.

\begin{figure}[]
    \centering
    \includegraphics[width=\columnwidth]{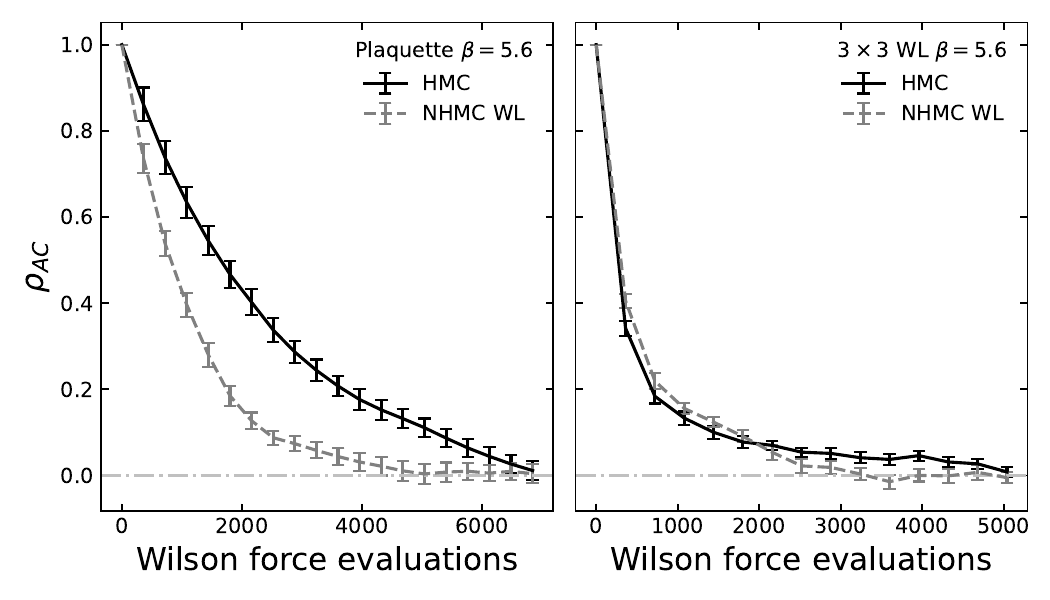}
    \captionsetup{skip=0pt}
    \caption{Comparison at $\beta = 5.6$ of the normalized autocorrelation $\rho_{\text{AC}}$ decreasing as a function of the number of Wilson force evaluations for the HMC and Nambu HMC with $3\times 3$ Wilson loops (NHMC WL). The left and right plots are the autocorrelation of the plaquette and $3\times3$ Wilson loop, respectively.}
    \label{b56_linG_WL}
\end{figure}

Test two is performed at a stronger coupling $\beta = 5.6$, which is just below the confining phase transition. The results are presented in Fig.~\ref{b56_linG_WL}. This test has 360 Wilson force evaluations per trajectory and utilizes 5000 trajectories. For the HMC, this corresponds to a total trajectory length of $t=9.0$ MD time units with acceptance rate $0.785$ and for the Nambu HMC, a trajectory length of $t=5.85$ MD time units with acceptance rate $0.787$. We find that the Nambu HMC again produces more rapid plaquette decorrelation than the HMC; however, in this case, we also observe slightly more rapid decorrelation for the Wilson loop.

\begin{figure}[]
    \centering
    \includegraphics[width=\columnwidth]{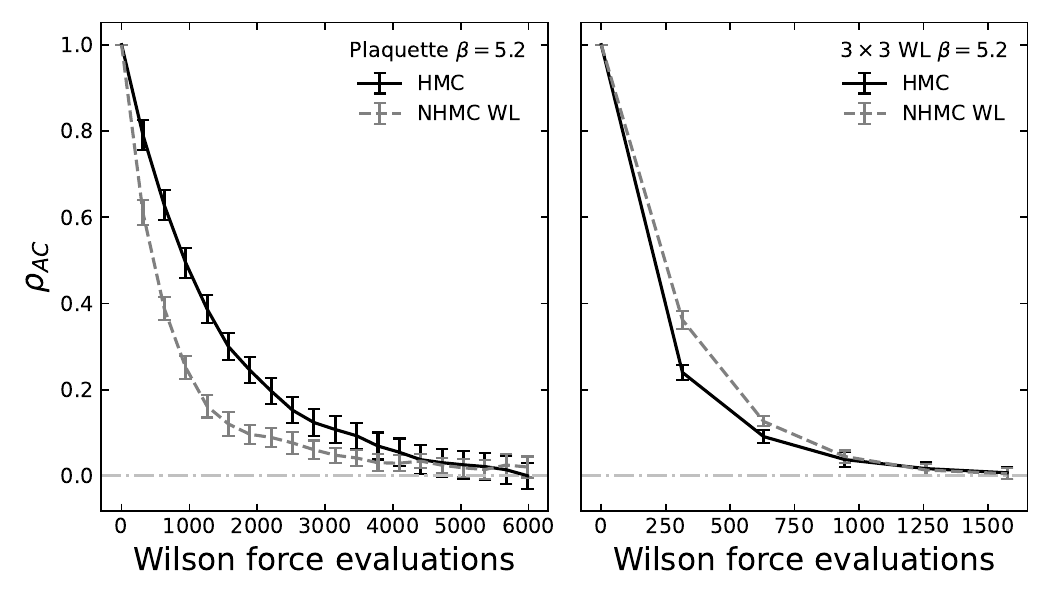}
    \captionsetup{skip=0pt}
    \caption{Comparison at $\beta = 5.2$ of the observable autocorrelation $\rho_{\text{AC}}$ plotted as a function of the number of trajectories for the HMC and Nambu HMC with $3\times 3$ Wilson loops (NHMC WL). The left and right plots are the autocorrelation of the plaquette and $3\times3$ Wilson loop, respectively.}
    \label{b52_linG_WL}
\end{figure}

Test three is performed at an even stronger coupling $\beta = 5.2$, which is well below the confining phase transition. The results are presented in Fig.~\ref{b52_linG_WL}. For this test there are 315 Wilson force evaluations per trajectory and we utilize 5000 trajectories. For the HMC, this corresponds to a total trajectory length of $t=9.9$ MD time units with acceptance rate $0.779$ and for the Nambu HMC, a total trajectory of length $t=6.75$ MD time units with acceptance rate $0.728$. Here again, we find a more rapid plaquette decorrelation, though we find a slower Wilson loop decorrelation.

\section{Conclusion}
\label{conclusion}

We have presented a novel generalization of the HMC algorithm that uses Nambu mechanics for molecular dynamics trajectories, along with a set of discrete updates suitable for lattice gauge theory. This mechanics is characterized by multiple Hamiltonian functions, only one of which is required for an accept/reject step. The remaining Hamiltonian can contain arbitrary non-local functions of the gauge field. This allows link variables to be updated with instantaneous knowledge of the gauge fields at large distances while preserving the target probability distribution, a feature that may help reduce CSD in lattice QCD simulations. In tests of pure SU(3) gauge theory, we demonstrated that the Nambu HMC produces results consistent with an exact algorithm. We also found that the auxiliary Hamiltonian can be used to improve the sampling efficiency of observables, as evidenced by the reduced autocorrelation of the plaquette. While these initial results do not come from carefully optimized comparisons on appropriately large lattice volumes, they suggest that a more carefully chosen auxiliary Hamiltonian could improve efficiency, motivating further exploration into optimizing this framework. 

This algorithm accommodates dynamical fermions in the same way as the standard HMC without requiring additional fermion force evaluations. Furthermore, the cost of evaluating forces from non-local observables in the auxiliary Hamiltonian is likely negligible compared to fermion forces. This makes the Nambu HMC an economical way to include communication across large distances in lattice QCD simulations. Along with adding fermions, the logical next step is to identify a set of non-local functions whose forces may be helpful in reducing critical slowing down in lattice QCD simulations, a direction that is currently under investigation. While here our focus has been on exploiting non-locality, there is incredible freedom in choosing the auxiliary Hamiltonian and it may be the case that a more exotic choice proves most beneficial.

\section{Acknowledgments}

Thank you to my Columbia collaborators for many ideas and discussions, particularly to Norman Christ for his helpful feedback and encouragement. This work was supported in part by the U.S. Department of Energy (DOE) Grant No. DE-SC0011941 and in part by the Exascale Computing Project (17-SC-20-SC), a collaborative effort of the U.S. Department of Energy Office of Science and the National Nuclear Security Administration.

\bibliography{ref.bib}

\begin{thebibliography}{8}%
\makeatletter
\providecommand \@ifxundefined [1]{%
 \@ifx{#1\undefined}
}%
\providecommand \@ifnum [1]{%
 \ifnum #1\expandafter \@firstoftwo
 \else \expandafter \@secondoftwo
 \fi
}%
\providecommand \@ifx [1]{%
 \ifx #1\expandafter \@firstoftwo
 \else \expandafter \@secondoftwo
 \fi
}%
\providecommand \natexlab [1]{#1}%
\providecommand \enquote  [1]{``#1''}%
\providecommand \bibnamefont  [1]{#1}%
\providecommand \bibfnamefont [1]{#1}%
\providecommand \citenamefont [1]{#1}%
\providecommand \href@noop [0]{\@secondoftwo}%
\providecommand \href [0]{\begingroup \@sanitize@url \@href}%
\providecommand \@href[1]{\@@startlink{#1}\@@href}%
\providecommand \@@href[1]{\endgroup#1\@@endlink}%
\providecommand \@sanitize@url [0]{\catcode `\\12\catcode `\$12\catcode `\&12\catcode `\#12\catcode `\^12\catcode `\_12\catcode `\%12\relax}%
\providecommand \@@startlink[1]{}%
\providecommand \@@endlink[0]{}%
\providecommand \url  [0]{\begingroup\@sanitize@url \@url }%
\providecommand \@url [1]{\endgroup\@href {#1}{\urlprefix }}%
\providecommand \urlprefix  [0]{URL }%
\providecommand \Eprint [0]{\href }%
\providecommand \doibase [0]{https://doi.org/}%
\providecommand \selectlanguage [0]{\@gobble}%
\providecommand \bibinfo  [0]{\@secondoftwo}%
\providecommand \bibfield  [0]{\@secondoftwo}%
\providecommand \translation [1]{[#1]}%
\providecommand \BibitemOpen [0]{}%
\providecommand \bibitemStop [0]{}%
\providecommand \bibitemNoStop [0]{.\EOS\space}%
\providecommand \EOS [0]{\spacefactor3000\relax}%
\providecommand \BibitemShut  [1]{\csname bibitem#1\endcsname}%
\let\auto@bib@innerbib\@empty
\bibitem [{\citenamefont {Duane}\ \emph {et~al.}(1987)\citenamefont {Duane}, \citenamefont {Kennedy}, \citenamefont {Pendleton},\ and\ \citenamefont {Roweth}}]{DUANE1987216}%
  \BibitemOpen
  \bibfield  {author} {\bibinfo {author} {\bibfnamefont {S.}~\bibnamefont {Duane}}, \bibinfo {author} {\bibfnamefont {A.}~\bibnamefont {Kennedy}}, \bibinfo {author} {\bibfnamefont {B.~J.}\ \bibnamefont {Pendleton}},\ and\ \bibinfo {author} {\bibfnamefont {D.}~\bibnamefont {Roweth}},\ }\bibfield  {title} {\bibinfo {title} {Hybrid monte carlo},\ }\href {https://doi.org/https://doi.org/10.1016/0370-2693(87)91197-X} {\bibfield  {journal} {\bibinfo  {journal} {Physics Letters B}\ }\textbf {\bibinfo {volume} {195}},\ \bibinfo {pages} {216} (\bibinfo {year} {1987})}\BibitemShut {NoStop}%
\bibitem [{\citenamefont {Davies}\ \emph {et~al.}(1990)\citenamefont {Davies}, \citenamefont {Batrouni}, \citenamefont {Katz}, \citenamefont {Kronfeld}, \citenamefont {Lepage}, \citenamefont {Rossi}, \citenamefont {Svetitsky},\ and\ \citenamefont {Wilson}}]{PhysRevD.41.1953}%
  \BibitemOpen
  \bibfield  {author} {\bibinfo {author} {\bibfnamefont {C.~T.~H.}\ \bibnamefont {Davies}}, \bibinfo {author} {\bibfnamefont {G.~G.}\ \bibnamefont {Batrouni}}, \bibinfo {author} {\bibfnamefont {G.~R.}\ \bibnamefont {Katz}}, \bibinfo {author} {\bibfnamefont {A.~S.}\ \bibnamefont {Kronfeld}}, \bibinfo {author} {\bibfnamefont {G.~P.}\ \bibnamefont {Lepage}}, \bibinfo {author} {\bibfnamefont {P.}~\bibnamefont {Rossi}}, \bibinfo {author} {\bibfnamefont {B.}~\bibnamefont {Svetitsky}},\ and\ \bibinfo {author} {\bibfnamefont {K.~G.}\ \bibnamefont {Wilson}},\ }\bibfield  {title} {\bibinfo {title} {Fourier acceleration in lattice gauge theories. iii. updating field configurations},\ }\href {https://doi.org/10.1103/PhysRevD.41.1953} {\bibfield  {journal} {\bibinfo  {journal} {Phys. Rev. D}\ }\textbf {\bibinfo {volume} {41}},\ \bibinfo {pages} {1953} (\bibinfo {year} {1990})}\BibitemShut {NoStop}%
\bibitem [{\citenamefont {Schaefer}\ \emph {et~al.}(2011)\citenamefont {Schaefer}, \citenamefont {Sommer},\ and\ \citenamefont {Virotta}}]{Schaefer_2011}%
  \BibitemOpen
  \bibfield  {author} {\bibinfo {author} {\bibfnamefont {S.}~\bibnamefont {Schaefer}}, \bibinfo {author} {\bibfnamefont {R.}~\bibnamefont {Sommer}},\ and\ \bibinfo {author} {\bibfnamefont {F.}~\bibnamefont {Virotta}},\ }\bibfield  {title} {\bibinfo {title} {Critical slowing down and error analysis in lattice qcd simulations},\ }\href {https://doi.org/10.1016/j.nuclphysb.2010.11.020} {\bibfield  {journal} {\bibinfo  {journal} {Nuclear Physics B}\ }\textbf {\bibinfo {volume} {845}},\ \bibinfo {pages} {93–119} (\bibinfo {year} {2011})}\BibitemShut {NoStop}%
\bibitem [{\citenamefont {Nambu}(1973)}]{nambu_mech}%
  \BibitemOpen
  \bibfield  {author} {\bibinfo {author} {\bibfnamefont {Y.}~\bibnamefont {Nambu}},\ }\bibfield  {title} {\bibinfo {title} {Generalized hamiltonian dynamics},\ }\href {https://doi.org/10.1103/PhysRevD.7.2405} {\bibfield  {journal} {\bibinfo  {journal} {Phys. Rev. D}\ }\textbf {\bibinfo {volume} {7}},\ \bibinfo {pages} {2405} (\bibinfo {year} {1973})}\BibitemShut {NoStop}%
\bibitem [{\citenamefont {Duane}\ \emph {et~al.}(1986)\citenamefont {Duane}, \citenamefont {Kenway}, \citenamefont {Pendleton},\ and\ \citenamefont {Roweth}}]{DUANE1986143}%
  \BibitemOpen
  \bibfield  {author} {\bibinfo {author} {\bibfnamefont {S.}~\bibnamefont {Duane}}, \bibinfo {author} {\bibfnamefont {R.}~\bibnamefont {Kenway}}, \bibinfo {author} {\bibfnamefont {B.~J.}\ \bibnamefont {Pendleton}},\ and\ \bibinfo {author} {\bibfnamefont {D.}~\bibnamefont {Roweth}},\ }\bibfield  {title} {\bibinfo {title} {Acceleration of gauge field dynamics},\ }\href {https://doi.org/https://doi.org/10.1016/0370-2693(86)90940-8} {\bibfield  {journal} {\bibinfo  {journal} {Physics Letters B}\ }\textbf {\bibinfo {volume} {176}},\ \bibinfo {pages} {143} (\bibinfo {year} {1986})}\BibitemShut {NoStop}%
\bibitem [{\citenamefont {Callaway}\ and\ \citenamefont {Rahman}(1982)}]{PhysRevLett.49.613}%
  \BibitemOpen
  \bibfield  {author} {\bibinfo {author} {\bibfnamefont {D.~J.~E.}\ \bibnamefont {Callaway}}\ and\ \bibinfo {author} {\bibfnamefont {A.}~\bibnamefont {Rahman}},\ }\bibfield  {title} {\bibinfo {title} {Microcanonical ensemble formulation of lattice gauge theory},\ }\href {https://doi.org/10.1103/PhysRevLett.49.613} {\bibfield  {journal} {\bibinfo  {journal} {Phys. Rev. Lett.}\ }\textbf {\bibinfo {volume} {49}},\ \bibinfo {pages} {613} (\bibinfo {year} {1982})}\BibitemShut {NoStop}%
\bibitem [{\citenamefont {Kennedy}\ \emph {et~al.}(2013)\citenamefont {Kennedy}, \citenamefont {Silva},\ and\ \citenamefont {Clark}}]{Kennedy_2013}%
  \BibitemOpen
  \bibfield  {author} {\bibinfo {author} {\bibfnamefont {A.~D.}\ \bibnamefont {Kennedy}}, \bibinfo {author} {\bibfnamefont {P.~J.}\ \bibnamefont {Silva}},\ and\ \bibinfo {author} {\bibfnamefont {M.~A.}\ \bibnamefont {Clark}},\ }\bibfield  {title} {\bibinfo {title} {Shadow hamiltonians, poisson brackets, and gauge theories},\ }\bibfield  {journal} {\bibinfo  {journal} {Physical Review D}\ }\textbf {\bibinfo {volume} {87}},\ \href {https://doi.org/10.1103/physrevd.87.034511} {10.1103/physrevd.87.034511} (\bibinfo {year} {2013})\BibitemShut {NoStop}%
\bibitem [{\citenamefont {Leimkuhler}\ and\ \citenamefont {Reich}(2005)}]{leimkuhler_reich_2005}%
  \BibitemOpen
  \bibfield  {author} {\bibinfo {author} {\bibfnamefont {B.}~\bibnamefont {Leimkuhler}}\ and\ \bibinfo {author} {\bibfnamefont {S.}~\bibnamefont {Reich}},\ }\href {https://doi.org/10.1017/CBO9780511614118} {\emph {\bibinfo {title} {Simulating Hamiltonian Dynamics}}},\ Cambridge Monographs on Applied and Computational Mathematics\ (\bibinfo  {publisher} {Cambridge University Press},\ \bibinfo {year} {2005})\BibitemShut {NoStop}%
\end{thebibliography}%
\onecolumngrid


\appendix
\section{Proof of detailed balance}
\label{detailed_balance}

This section proves that an extended HMC algorithm using Nambu mechanics for MD steps satisfies the detailed balance condition
\begin{equation}
    P(U) P_T(U \rightarrow U') = P(U') P_T(U' \rightarrow U).
\label{DB_condition}
\end{equation}
\noindent For the purposes of this proof the Hamiltonian $H(p,U,r)$ to be used in the Metropolis accept/reject step is as in Eq.~\eqref{separable hamiltonians}
\begin{equation}
    H(p,U,r) = \frac{p^2}{2} + \frac{r^2}{2} + S(U).
\label{H_DB}
\end{equation}
\noindent The form of the auxiliary Hamiltonian $G$ is the same as in the main text. The proof proceeds nearly identically to the one used for the usual HMC \cite{DUANE1987216}. At the beginning of each trajectory, the fictitious variables $p$ and $r$ are generated at random from a Gaussian distribution $P_G(p)$ and $P_G(r)$ where
\begin{equation}
    P_G(z) \propto \exp(-\frac{1}{2} z^2).
\label{p_g}
\end{equation}
\noindent The MD trajectories are simulated for a total computer time $t$ and at the end of a trajectory one performs a Metropolis accept/reject test with acceptance probability min$(1, e^{-\Delta H})$.  

Evolution via Nambu's evolution equations for time $t$ is a map in phase space $(p(0),U(0),r(0)) \rightarrow (p(t),U(t),r(t))$. The probability for choosing a particular point in phase space $(p',U',r')$ is 
\begin{equation}
    P_H\left[(p,U,r) \rightarrow (p',U',r')\right] = \delta \left[(p',U',r') - (p(t),U(t),r(t))\right].
\label{p_h}
\end{equation}
\noindent The probability of accepting this change is given by
\begin{equation}
    P_A \left[(p,U,r) \rightarrow (p',U',r')\right] = \text{min}(1, e^{-\Delta H}),
\label{p_a}
\end{equation}
\noindent where $\Delta H = H(p',U',r') - H(p,U,r)$. Together, Eqs. (\ref{p_g}-\ref{p_a}) yield a probability of transition for the gauge field of
\begin{equation}
    P_T(U \rightarrow U') = \int [dp][dp'][dr][dr'] P_G(p) P_G(r) P_H \left[ (p,U,r) \rightarrow (p',U',r') \right] P_A \left[(p,U,r) \rightarrow (p',U',r')\right].
\label{gauge_field_transition_prob}
\end{equation}
\noindent A necessary condition for detailed balance is that the evolution be reversible. As explained in Section \ref{nambu_mech}, with Hamiltonians satisfying $H(-p,U,r) = H(p,U,r)$ and $G(-p,U,r)=G(p,U,r)$, the reverse trajectory is obtained by taking $p \rightarrow -p$ so that
\begin{equation}
    P_H\left[(p,U,r) \rightarrow (p',U',r')\right] = P_H \left[(-p',U',r') \rightarrow (-p,U,r)\right].
\end{equation}
\noindent Given the properties
\begin{equation}
\exp(-H(p,U,r))\text{min}(1, e^{-\Delta H}) = \exp(-H(p',U',r'))\text{min}(e^{\Delta H},1),
\end{equation}
\noindent and
\begin{equation}
    P_G(p) P_G(r) P(U) \propto \exp(-H(p,U,r)),
\end{equation}
\noindent the following is true
\begin{equation}
\begin{aligned}
    P_G(p) P_G(r) P(U) P_A \left[(U,p,r) \rightarrow (p',U',r')\right] & = P_G(p') P_G(r') P(U') P_A \left[(p',U',r') \rightarrow (p,U,r)\right] \\
    & = P_G(-p') P_G(r') P(U') P_A \left[(-p',U',r') \rightarrow (-p,U,r)\right].
\end{aligned}
\end{equation}
\noindent Multiplying by $P_H$ and integrating over the fictitious momenta one finds
\begin{equation}
\begin{aligned}
    \int [dp][dp'][dr][dr'] & P(U) P_G(p) P_G(r) P_H \left[(p,U,r) \rightarrow (p',U',r')\right] P_A \left[(p,U,r) \rightarrow (p',U',r')\right] \\
     & = \int [d(-p)][d(-p')][dr][dr'] P(U') P_G(-p') P_G(r') P_H \left[(-p',U',r') \rightarrow (-p,U,r)\right] P_A \left[(-p',U',r') \rightarrow (-p,U,r)\right].
\end{aligned}
\end{equation}
\noindent Considering the invariance of the measure $[dp][dp']= [d(-p)][d(-p')]$, this is the detailed balance condition in Eq.~\eqref{DB_condition}).

\end{document}